\documentclass[conference]{IEEEtran}
\IEEEoverridecommandlockouts

\usepackage{textcomp}
\usepackage{xcolor}

\usepackage{amsmath,amssymb,amsfonts}
\usepackage{algorithm}
\usepackage{algorithmic}
\usepackage{graphicx}
\usepackage{tabularx}
\graphicspath{ {Figures/} }
\usepackage{stfloats}
\usepackage{multicol}
\usepackage{multirow}
\usepackage{tablefootnote}
\usepackage[flushleft]{threeparttable}
\usepackage{color, colortbl}
\usepackage{lipsum}

\usepackage{hyperref}

\usepackage{balance}

\usepackage[noadjust]{cite}

\newcommand{\Fp}{\mathbb{F}_p}
\newcommand{\K}{\mathbb{K}}

\definecolor{Gray}{gray}{0.9}
\definecolor{Highlight}{rgb}{0.99, 0.99, 0.9}

\def\BibTeX{{\rm B\kern-.05em{\sc i\kern-.025em b}\kern-.08em
    T\kern-.1667em\lower.7ex\hbox{E}\kern-.125emX}}
\begin{document}

\title{
A High-Performance Curve25519 and Curve448 Unified Elliptic Curve Cryptography Accelerator
\thanks{\textcopyright $\,$ 2024 IEEE. Personal use of this material is permitted. Permission from IEEE must be obtained for all other uses, in any current or future media, including reprinting/republishing this material for advertising or promotional purposes, creating new collective works, for resale or redistribution to servers or lists, or reuse of any copyrighted component of this work in other works.}
\thanks{A revised version of this paper was published in the proceedings of the 2024 IEEE High Performance Extreme Computing Conference (HPEC) - DOI: \href{https://dx.doi.org/10.1109/HPEC62836.2024.10938523}{10.1109/HPEC62836.2024.10938523}}
}

\author{
\IEEEauthorblockN{Aniket Banerjee and Utsav Banerjee}
\IEEEauthorblockA{Electronic Systems Engineering \\
Indian Institute of Science, Bengaluru, India \\
Email: aniketb@iisc.ac.in, utsav@iisc.ac.in}
}

\maketitle

\begin{abstract}
In modern critical infrastructure such as power grids, it is crucial to ensure security of data communications between network-connected devices while following strict latency criteria. This necessitates the use of cryptographic hardware accelerators.
We propose a high-performance unified elliptic curve cryptography accelerator supporting NIST standard Montgomery curves Curve25519 and Curve448 at 128-bit and 224-bit security levels respectively.
Our accelerator implements extensive parallel processing of Karatsuba-style large-integer multiplications, restructures arithmetic operations in the Montgomery Ladder and exploits special mathematical properties of the underlying pseudo-Mersenne and Solinas prime fields for optimized performance. Our design ensures efficient resource sharing across both curve computations and also incorporates several standard side-channel countermeasures.
Our ASIC implementation achieves record performance and energy of 10.38~$\mu$s / 54.01~$\mu$s and 0.72~$\mu$J / 3.73 $\mu$J respectively for Curve25519 / Curve448, which is significantly better than state-of-the-art.
\end{abstract}


\section{Introduction}
\label{sec:introduction}

The need for enhanced cybersecurity is more critical than ever in today's interconnected world. As digital communications and transactions become ubiquitous, ensuring information confidentiality, integrity and authenticity has become a paramount concern. Public Key Cryptography (PKC) \cite{paar_crypto_2009, menezes_handbook_2018} is pivotal in securing these communications by enabling key establishment, digital signatures and authentication protocols. This is especially important in the Internet of Things (IoT) such as industrial automation, sensors, power grids, smart cities and automotive applications \cite{keoh_iotsurvey_2014}. As these sectors continue to digitize, ensuring robust security while maintaining operational efficiency becomes a significant challenge. For example, digital communications between intelligent electronic devices in modern electrical substations follow the IEC 61850 standard \cite{iec61850_standard, mackiewicz_overview_2006} and their security recommendations are provided by the IEC 62351 standard \cite{iec62351_standard, hussain_review_2020}. These protocols demand low latency and high reliability for associated messaging protocols for real-time operation, e.g., 3~ms and 250~$\mu$s for GOOSE (Generic Object Oriented Substation Event) and Sampled Value (SV) messages respectively, thus making it extremely challenging to implement strong cryptographic security measures in critical infrastructure \cite{hohlbaum_practical_2010}. Due to the use of embedded systems, it is also crucial to ensure hardware resilience against side-channel attacks \cite{verbauwhede_embedded_2006}.

Cryptographic hardware accelerators are widely used to meet application-specific requirements such as low power, high performance and energy efficiency which are not achievable using software implementations with general purpose micro-processors \cite{banerjee_phd_2021}.
Elliptic Curve Cryptography (ECC) is the current standard for PKC algorithms due to small key sizes \cite{koblitz_ecc_1987, hankerson_ecc_2006}, and the U.S. National Institute of Standards and Technology (NIST) has recently recommended two new elliptic curves Curve25519 and Curve448 \cite{nist_ecparams_2023}. 
Both Curve25519 and Curve448 are Montgomery curves which stand out for their exceptional performance and security properties, as specified in \cite{ietf_rfc7748_2016}.
Curve25519, introduced by Daniel J. Bernstein in 2006 \cite{bernstein_curve25519_2006}, is renowned for its speed and robustness against side-channel attacks. Curve448, introduced by Mike Hamburg in 2015 \cite{hamburg_curve448_2015}, offers a higher security level while maintaining efficiency and side-channel resilience.
In spite of the advent of new quantum-safe cryptography algorithms, elliptic curves such as Curve25519 and Curve448 continue to play a vital role in enabling post-quantum hybrid key exchange protocols due to the strong confidence in their security \cite{giron_post_2023}.

Previous literature has presented various dedicated hardware accelerator designs for Curve25519 and Curve448 implemented in FPGA (field-programmable gate array) and ASIC (application-specific integrated circuit) \cite{
sasdrich_efficient_2014,
sasdrich_implementing_2015,
hutter_nacl_2015,
sasdrich_closing_2016,
koppermann_x25519_2016,
sasdrich_cryptography_2017,
koppermann_low_2017,
sasdrich_exploring_2018,
turan_compact_2019,
mehrabi_low_2019,
salarifard_efficient_2019,
shah_lut_2020,
niasar_optimized_2020,
niasar_fast_2020,
yang_hardware_2021,
bisheh_cryptographic_2021,
yu_high_2021,
bisheh_area_2021,
mondal_hardware_2021,
kieu_low_2022,
awaludin_high_2022,
wu_high_2022,
do_multi_2023}.
Despite the numerous similarities between Curve25519 and Curve448, a unified hardware architecture supporting ECC with both curves is yet to be explored.
In this work, we present the ASIC implementation of a high-performance unified hardware accelerator which can be configured to perform elliptic curve scalar multiplication over both Curve25519 and Curve448. Since finite field arithmetic is the most expensive component of these computations, we propose an efficient modular arithmetic architecture with four 256-bit 2-level Karatsuba multipliers. This allows us to exploit the power of parallel processing for Curve25519, while the same can also be fully re-used for Curve448 by exploiting the special structure and mathematical properties of its underlying prime field. We re-structure and re-arrange the sequence of operations in the Montgomery Ladder to reduce the latency. Our design is constant-time by design and also incorporates the randomized projective coordinate countermeasure against power side-channel attacks.
Compared to previous designs, our proposed accelerator excels in terms of both performance and energy-efficiency while supporting a higher security level along with side-channel countermeasures.
\section{Background}
\label{sec:background}

\subsection{Elliptic Curve Cryptography (ECC)}

An elliptic curve $E$ over a finite field $\K$ is defined as
\[
E: y^2 + a_1xy + a_3y = x^3 + a_2x^2 + a_4x + a_6
\]
where $a_1, a_2, a_3, a_4, a_6 \in \K$.
There are two major types of elliptic curves defined over finite fields where the characteristic char($\K$) is a very large prime $p$:
\begin{itemize}
\item Short Weierstrass curves consisting of the set of points $E(\Fp)$ = \{($x$, $y$) \big| $y^2 = x^3 + ax + b$ (mod $p$)\} $\cup$ $\mathcal O$
\item Montgomery curves consisting of the set of points $E(\Fp)$ = \{($x$, $y$) \big| $by^2 = x^3 + ax^2 + x$ (mod $p$)\} $\cup$ $\mathcal O$
\end{itemize}
where $a, b \in \Fp$ are the curve parameters and $\mathcal O$ is the distinguished point at infinity.

The fundamental operations in ECC are \textit{point addition} ($R = P + Q$) and \textit{point doubling} ($R = P + P$), where $P, Q, R \in E(\Fp)$. With these operations, the points on the curve $E(\Fp)$ form an abelian group, with $\mathcal O$ serving as the identity element, that is, $P + \mathcal O = \mathcal O + P = P$ for all $P \in E(\Fp)$. The order of this group (number of points in $E(\Fp)$) is denoted by $\#E(\Fp) = n$, and $nP = \mathcal O$ for all $P \in E(\Fp)$.

Repeated additions of a point $P$ with itself is called \textit{elliptic curve scalar multiplication} (ECSM). For any scalar $k$, the scalar multiple $kP$ is computed as
\[
\underbrace{kP = P + P + \cdots + P}_{\text{($k$-1) point additions}}
\]
This computation forms the basis of the \textit{elliptic curve discrete logarithm problem} (ECDLP) -- determine scalar $k$ given the elliptic curve $E(\Fp)$ of order $n$, and the points $P, Q \in E(\Fp)$ such that $Q = kP$. For a $t$-bit prime $p$, the fastest known algorithms that can solve ECDLP have time complexity $O(2^{t/2})$ \cite{hankerson_ecc_2006}. For sufficiently large primes and appropriate curve parameters, it is infeasible for a computationally bounded (non-quantum) adversary to solve ECDLP, and this guarantees the security of ECC and associated public key protocols.

\subsection{Curve25519}

Curve25519 is a Montgomery curve defined by the equation \[y^2 = x^3 + 486662x^2 + x\] over the prime field \(\mathbb{F}_{2^{255}-19}\) \cite{bernstein_curve25519_2006}. It is optimized for key exchange at the 128-bit security level and has smaller key sizes and faster computations compared to traditional elliptic curves. It is designed to resist side-channel attacks, making it suitable for a wide range of applications including secure communications. Its simplicity and efficiency have led to its widespread adoption in various standards \cite{nist_ecparams_2023, ietf_rfc7748_2016}.

\subsection{Curve448}

Curve448 is a Montgomery curve defined by the equation \[y^2 = x^3 + 156326x^2 + x\] over the prime field \(\mathbb{F}_{2^{448}-{2^{224}-1}}\) \cite{hamburg_curve448_2015}. It offers a higher 224-bit security level while maintaining similar characteristics as Curve25519. It is optimized for cryptographic protocols requiring robust security such as key exchange and digital signatures. Its resistance to side-channel attacks makes it suitable for high-security applications \cite{nist_ecparams_2023, ietf_rfc7748_2016}.

\subsection{Hardware Acceleration of Curve25519 and Curve448}

Previous work on hardware implementations of Curve25519 have been mostly based on Zynq 7000 series FPGAs \cite{sasdrich_efficient_2014, sasdrich_implementing_2015, koppermann_x25519_2016, koppermann_low_2017, sasdrich_exploring_2018, turan_compact_2019, mehrabi_low_2019, niasar_fast_2020, mondal_hardware_2021, yang_hardware_2021, bisheh_cryptographic_2021, yu_high_2021, kieu_low_2022}.
The general approach has been to efficiently utilize the on-chip DSP and BRAM slices to speed up the ECSM operation. Various side-channel countermeasures such as scalar blinding, randomized projective coordinates and memory address scrambling have also been proposed \cite{fan_ecc_2010}.
Most recently, \cite{wu_high_2022} demonstrated a high-performance design consisting of compute groups with processing elements (PEs), massive parallelism and a high degree of pipelining implemented in a Zynq 7000 series FPGA.
The first compact low-power ASIC implementation of Curve25519 was presented in \cite{hutter_nacl_2015}, and this was subsequently optimized for high-performance in \cite{salarifard_efficient_2019, do_multi_2023}.
A similar approach was followed for FPGA-based hardware implementations of Curve448 \cite{sasdrich_closing_2016, sasdrich_cryptography_2017, shah_lut_2020, bisheh_area_2021, awaludin_high_2022}, with ECSM performance significantly slower than Curve25519 due to the increased computational complexity. Various implementation strategies for Curve448 hardware architectures in FPGA, such as light-weight, area-time-efficient and high-performance, were investigated by \cite{niasar_optimized_2020}. Efficient ASIC implementations of Curve448 are yet to be explored. Also, unified hardware architectures for accelerating ECSM over Curve25519 and Curve448 have not yet been demonstrated in state-of-the-art, thus motivating our proposed design in this work.
\section{Accelerator Architecture}
\label{sec:architecture}

\subsection{ECSM Computation using Montgomery Ladder}

Curve25519 and Curve448 are elliptic curves offering 128-bit and 224-bit of security levels respectively. Both curves use the Montgomery form, enabling fast and constant-time elliptic curve scalar multiplication, which is crucial for secure implementation. Supporting both curves in the same hardware accelerator facilitates interoperability and flexibility, allowing systems to choose the appropriate curve based on application-specific security needs. The proposed unified implementation also benefits from sharing hardware resources for the core arithmetic computations, thus reducing overall power and area without compromising performance.

\begin{algorithm}[!t]
\caption{ECSM using the Montgomery Ladder \cite{joye_ladder_2002}}
\label{algo:main}
\begin{algorithmic}[1]
\REQUIRE input point $P$ with $x$-coordinate $x_P$ and $t$-bit secret scalar $k = (k_{t-1}, k_{t-2}, \cdots, k_2, k_1, k_0)_2$
\ENSURE output point $Q = kP$ with $x$-coordinate $x_Q$
\STATE $X_1 \leftarrow x_P$, $X_2 \leftarrow 1$, $X_3 \leftarrow x_P$
\STATE $Z_1 \leftarrow 1$, $Z_2 \leftarrow 0$, $Z_3 \leftarrow 1$
\FOR{($i = t-1$; $i \ge 0$; $i = i - 1$)}
\IF{$k_i = 1$}
\STATE $(X_3, Z_3, X_2, Z_2) \leftarrow \text{LADDER} \, (X_1, X_3, Z_3, X_2, Z_2)$
\ELSE
\STATE $(X_2, Z_2, X_3, Z_3) \leftarrow \text{LADDER} \, (X_1, X_2, Z_2, X_3, Z_3)$
\ENDIF
\ENDFOR
\STATE $Z_2 \leftarrow Z_2^{-1}$
\STATE $x_Q \leftarrow X_2 \, Z_2$
\RETURN $x_Q$
\end{algorithmic}
\end{algorithm}

\begin{figure} [!t]
\centering
\includegraphics[width=3.2in]{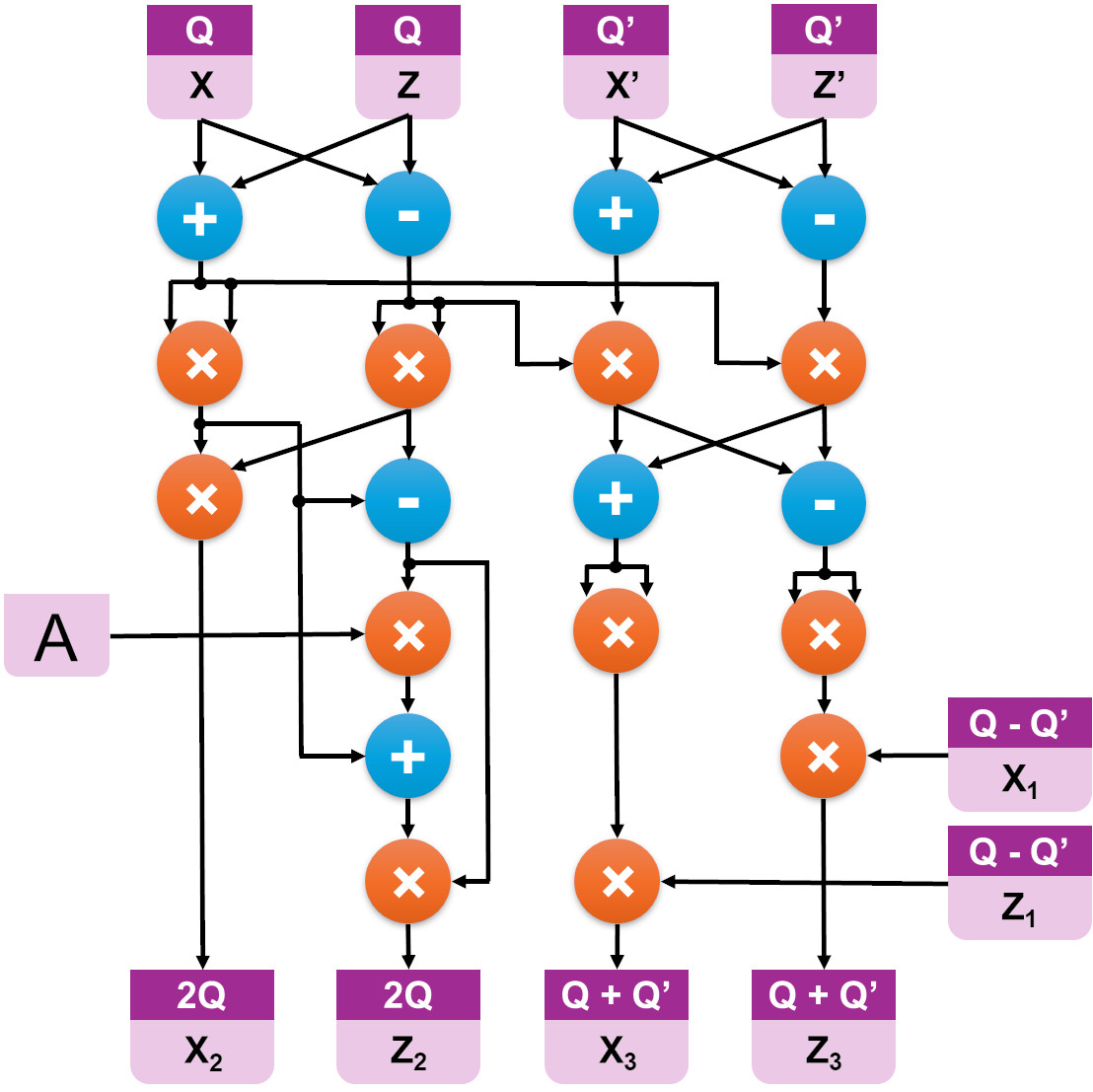}
\caption{Modular arithmetic operations in the Montgomery Ladder.}
\label{fig:ladder}
\end{figure}

Algorithm \ref{algo:main} shows how an ECSM computation is performed on a Montgomery curve with $t$-bit scalar $k = (k_{t-1}, k_{t-2}, \cdots, k_2, k_1, k_0)_2$. Note that $t = 255$ and $t=448$ for Curve25519 and Curve448, respectively. The input and output points $P$ and $Q$ are specified by their $x$-coordinates $x_P$ and $x_Q$, respectively. For Curve25519 and Curve448, $x_P$ and $x_Q$ will be elements in their respective prime fields $\mathbb{F}_{2^{255}-19}$ and $\mathbb{F}_{2^{448}-{2^{224}-1}}$.
This algorithm is inherently constant-time, that is, execution time is independent of the secret scalar $k$. This helps prevent timing and simple power analysis (SPA) attacks. In order to prevent more sophisticated differential power analysis (DPA) attacks, the randomized projective coordinate technique can be used \cite{fan_ecc_2010}. This involves first generating a pseudo-random element $\lambda$ in the underlying field ($\mathbb{F}_{2^{255}-19}$ for Curve25519 and $\mathbb{F}_{2^{448}-{2^{224}-1}}$ for Curve448). Then, steps 1 and 2 in Algorithm \ref{algo:main} are modified as $X_1 \leftarrow \lambda \, x_P$, $X_2 \leftarrow \lambda$, $X_3 \leftarrow \lambda \, x_P$ and $Z_1 \leftarrow \lambda$, $Z_2 \leftarrow 0$, $Z_3 \leftarrow \lambda$ respectively. Rest of the ECSM computation remains unchanged, and the modular division in steps 10 and 11 in Algorithm \ref{algo:main} ensure that the final output is correct irrespective of the value of $\lambda$.

\begin{figure}[!t]
\centering
\includegraphics[width=3.4in]{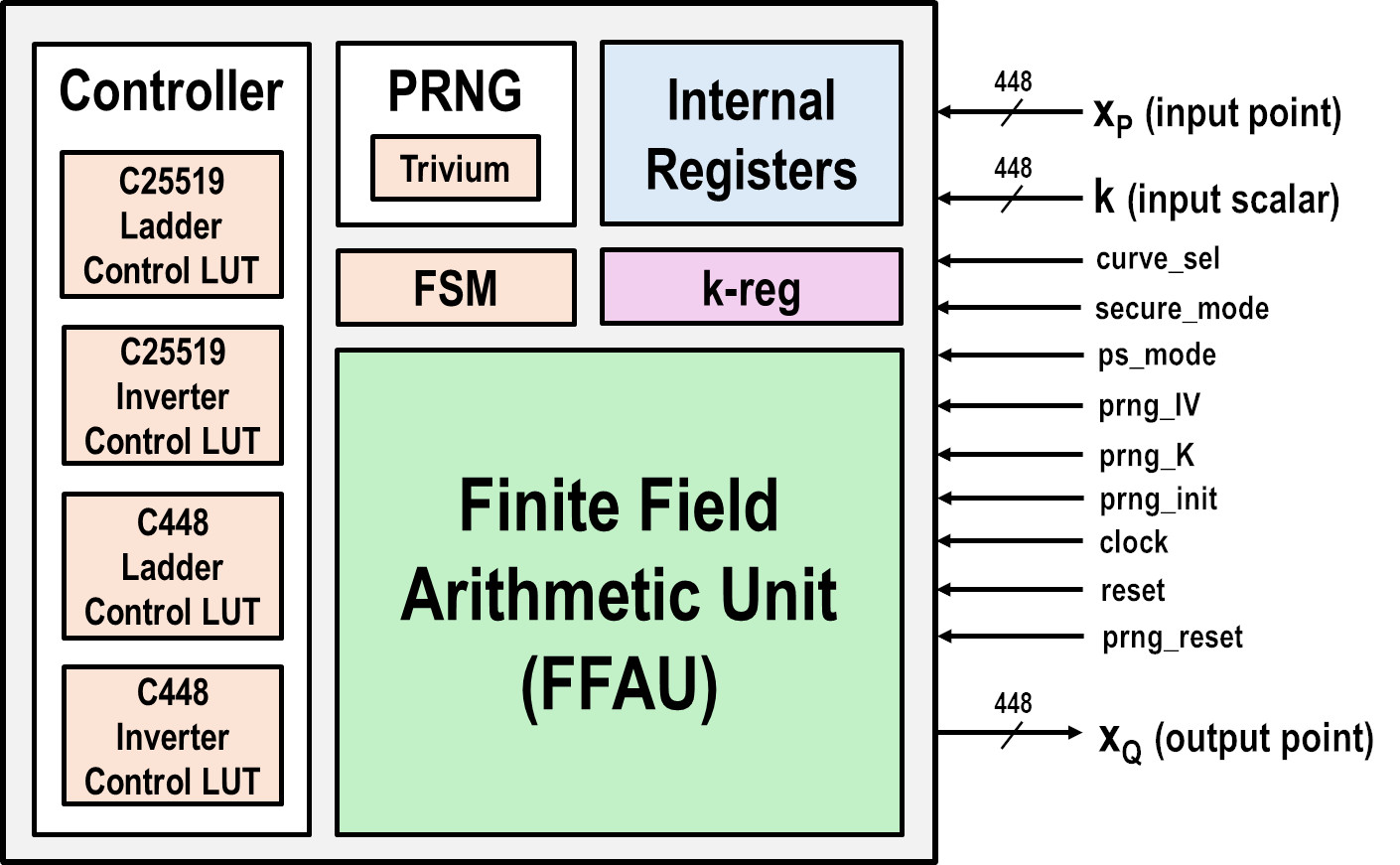}
\caption{Top-level block diagram of the proposed unified Curve25519 and Curve448 elliptic curve cryptography accelerator.}
\label{fig:top_module}
\end{figure}

\begin{figure}[!t]
\centering
\includegraphics[width=3.4in]{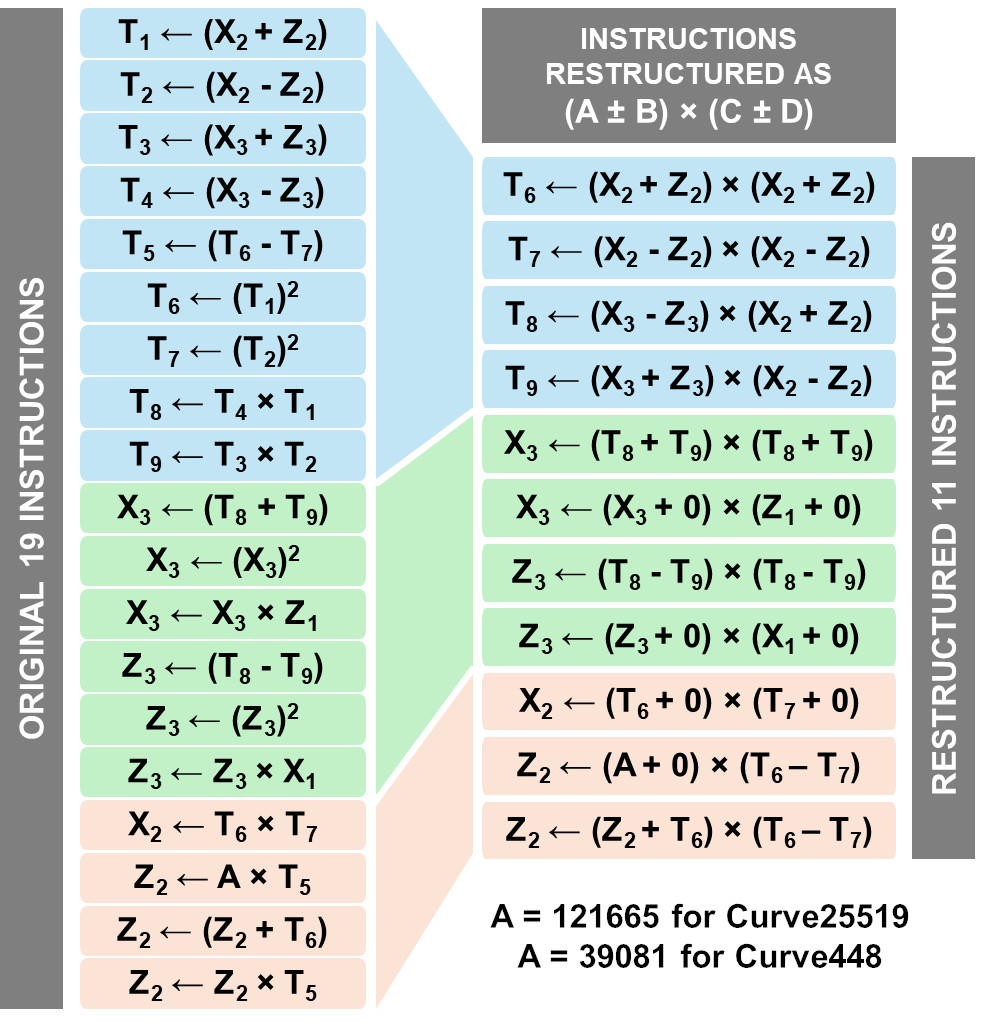}
\caption{Restructuring of arithmetic operations in the LADDER computation.}
\label{fig:ladder_ins}
\end{figure}

The most important step in Algorithm \ref{algo:main} is the LADDER(.) function, which is the Montgomery Ladder \cite{joye_ladder_2002}.
Both Curve25519 and Curve448 ECSM computations employ the Montgomery Ladder to perform point double-and-add operations in projective coordinates, and its constituent modular arithmetic operations are shown in Fig. \ref{fig:ladder}.
There are 8 modular additions/subtractions and 11 modular multiplications/squarings (including multiplication by the curve constant $A$, where $A = 121665$ for Curve25519 and $A = 39081$ for Curve448) involved in each LADDER computation.

\subsection{Accelerator Building Blocks}

The top-level block diagram of our proposed hardware accelerator is shown in Fig. \ref{fig:top_module}.
The most important component of the accelerator is the Finite Field Arithmetic Unit (FFAU).
A 448-bit k-reg register is used to store the secret scalar $k$.
A unified controller module is used to send appropriate control signals and instructions to the main data-path, and they work in tandem with a finite state machine (FSM).
12 $\times$ 448-bit internal registers are used to store the inputs, outputs and temporary values generated during ECSM computation. The most significant 193 bits of all these registers are clock-gated for power savings when performing ECSM over Curve25519, while all 448 bits are utilized for Curve448.
A pseudo-random number generator (PRNG) module, containing a hardware instantiation of the light-weight Trivium stream cipher \cite{preneel_trivium_2006}, is used to generate the $\lambda$ values for DPA countermeasures as discussed earlier.
The PRNG is initialized by the inputs prng\_IV and prng\_K.
The PRNG is clock-gated for power savings when DPA countermeasures are disabled.

\begin{figure} [!t]
\centering
\includegraphics[width=3.4in]{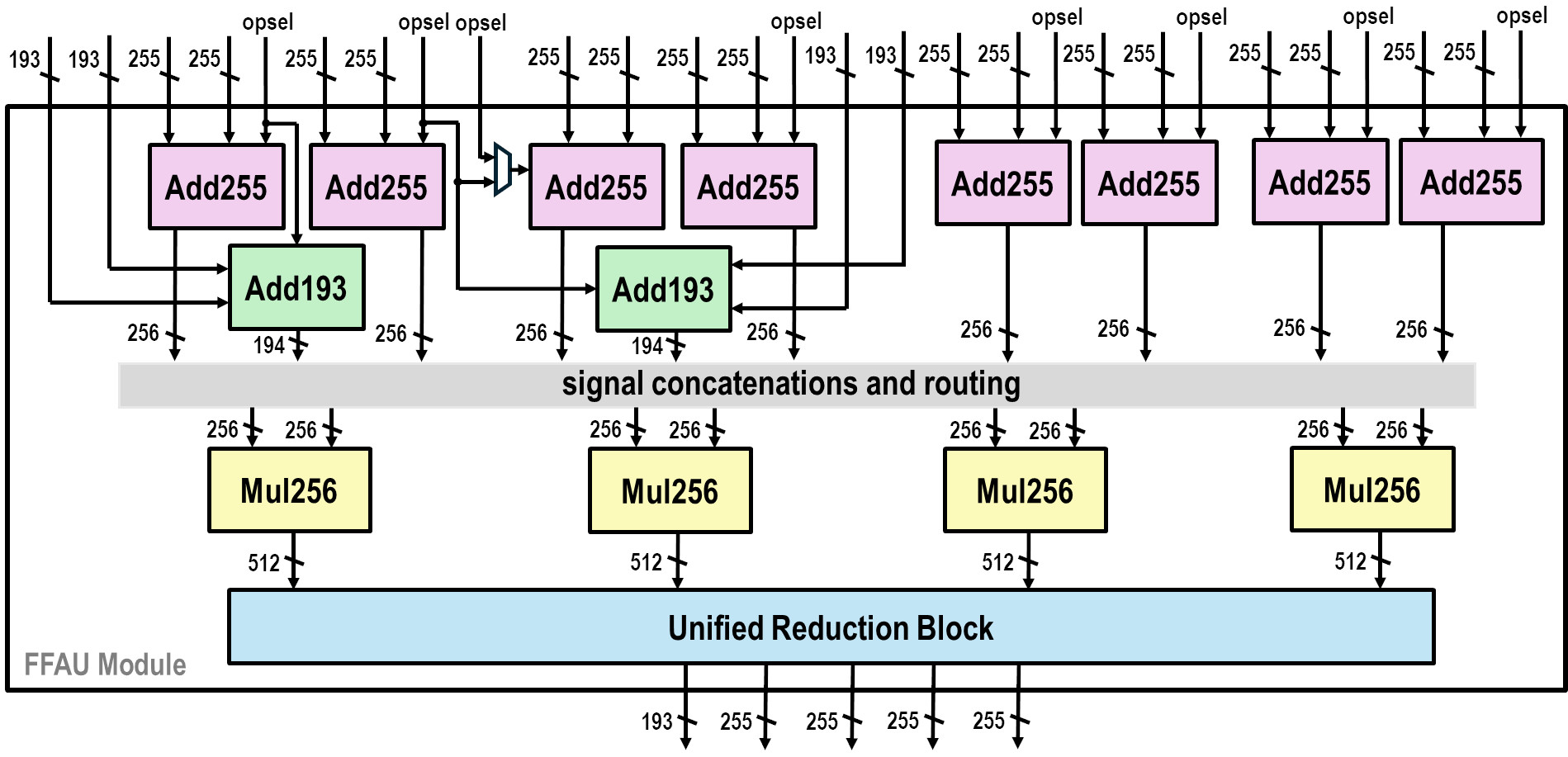}
\caption{Detailed architecture of the finite field arithmetic unit (FFAU) module.}
\label{fig:FFAU_top}
\end{figure}

\subsection{Unified Controller}

Inspired by the instruction mapping technique from \cite{wu_high_2022}, we restructure the 19 modular arithmetic operations in the LADDER computation from Fig. \ref{fig:ladder} as 11 steps in the form of \((A \pm B) \times (C \pm D)\). This restructuring is shown in Fig. \ref{fig:ladder_ins}.
Control signals and register addresses for these steps in the LADDER computation for both curves are stored as instructions in lookup tables (LUTs), as shown in Fig. \ref{fig:top_module}.
For ECSM computation, 255 and 448 iterations of the LADDER are performed for Curve25519 and Curve448, respectively.
The controller also contains similar LUTs for the modular inversion computations which will be discussed later.

\subsection{Finite Field Arithmetic Unit (FFAU)}

All the modular arithmetic operations in our accelerator are executed in the FFAU module shown in Fig. \ref{fig:FFAU_top}. It is capable of computing four \((A \pm B) \times (C \pm D)\) operations simultaneously, where $A$, $B$, $C$, $D$ are 255-bit inputs. The FFAU has eight \textit{opsel} (operation select) input lines corresponding to the four operations to select whether additions or subtractions need to be performed.
The FFAU contains four 256-bit multipliers \textit{Mul256} and eight 255-bit adder / subtractor modules \textit{Add255}. It also contains two 193-bit adder / subtractor modules \textit{Add193} to together compute 448-bit addition / subtraction.
Clearly, the FFAU can perform four instructions together for Curve25519, thus completing a LADDER in just 3 clock cycles.
When working with Curve448, the fact that its prime modulus is a Solinas trinomial prime ($= \phi^2 - \phi - 1$) with the golden ratio \(\phi = 2^{224}\) can be exploited. This allows the product of \(A = (a_{1}\phi + a_{0}) \in \mathbb{F}_{\phi^2 - \phi - 1}\) and \(B = (b_{1}\phi + b_{0}) \in \mathbb{F}_{\phi^2 - \phi - 1}\) to be calculated efficiently as: $C = A \times B \,\,\, (\text{mod} \,\,\, \phi^2 - \phi - 1) = (a_{1}\phi + a_{0}) \times (b_{1}\phi + b_{0}) \,\,\, (\text{mod} \,\,\, \phi^2 - \phi - 1) = (a_{1}b_{1} + a_{0}b_{0}) + (a_{1}b_{0} + a_{0}b_{1} + a_{0}b_{0}) \phi \,\,\, (\text{mod} \,\,\, \phi^2 - \phi - 1)$.
Here, $a_0$, $a_1$, $b_0$, $b_1$ being 224-bit quantities, it is possible to perform this computation using the four 256-bit multipliers. 
Therefore, the FFAU can perform only one instruction at a time for Curve448. Consequently, 10 clock cycles are required to complete a LADDER (11 clock cycles with DPA countermeasure enabled).
The multiplier outputs are finally processed by the \textit{Unified Reduction Block} which employs fast reduction algorithms to compute the final result.
Details of the FFAU sub-module implementations are described as follows:

\begin{figure}[!b]
\centering
\includegraphics[width=3.4in]{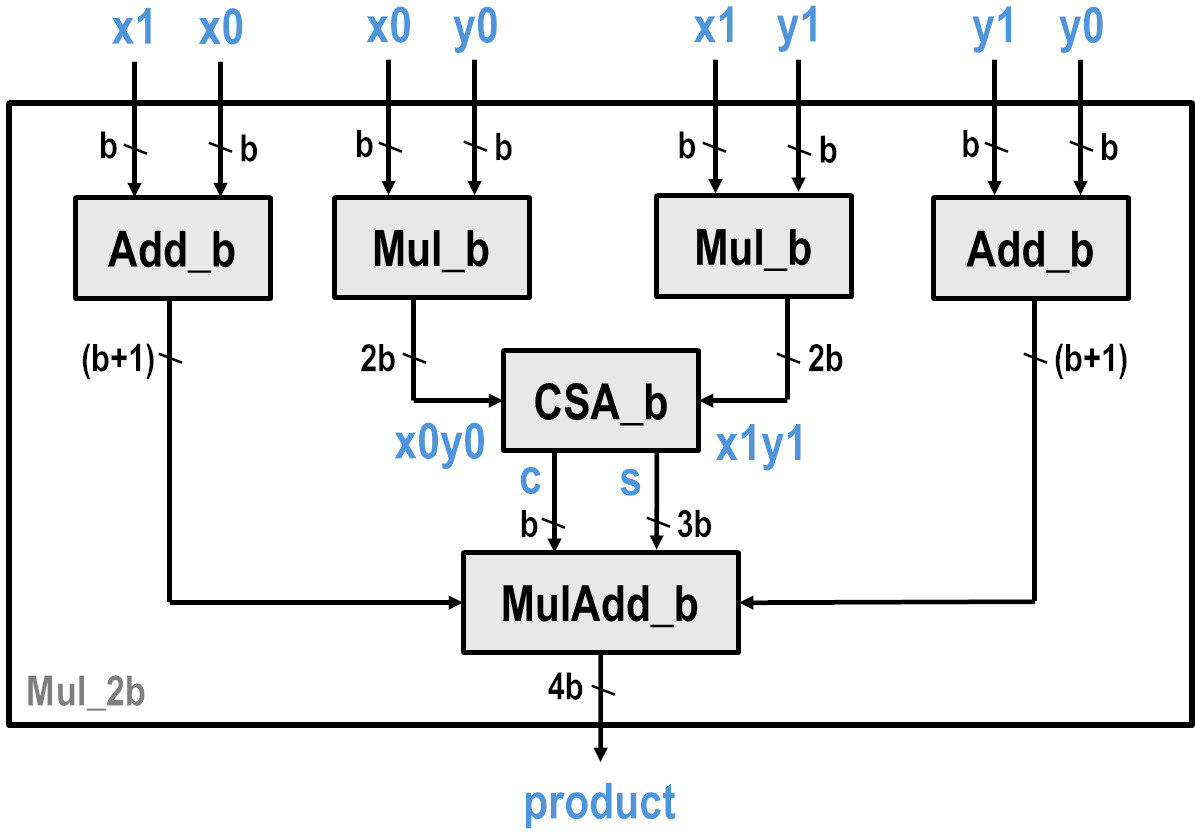}
\caption{Block diagram of $2b$-bit $\times$ $2b$-bit Karatsuba multiplier ($b = 128$ and $b = 64$ for Mul256 and Mul128 respectively).}
\label{fig:mult_b}
\end{figure}

\subsubsection{256-bit Multiplier}

The Mul256 modules are implemented as 256-bit 2-level Karatsuba multipliers \cite{karatsuba_multiplication_1962}. Using Karatsuba's algorithm, the product of two $2b$-bit unsigned integers $X = x_{1}2^{b} + x_{0}$ and $Y = y_{1}2^{b} + y_{0}$ can be calculated as $Z = XY = x_{1}y_{1}2^{2b} + (x_{0}y_{1} + x_{1}y_{0})2^{b} + x_{0}y_{0} = x_{1}y_{1}2^{2b} + [(x_{0}+x_{1})(y_{0}+y_{1}) - (x_{0}y_{0} + x_{1}y_{1})]2^{b} + x_{0}y_{0}$, that is, three instead of four $b$-bit $\times$ $b$-bit multiplications at the cost of some extra additions / subtractions. This reduces the complexity of $n$-bit large integer multiplication from \(O(n^{2})\) to \(O(n^{\log_2(3)})\), or approximately \(O(n^{1.585})\).
Fig. \ref{fig:mult_b} shows the block diagram for this construction. The Mul\_b blocks compute $x_{0}y_{0}$ and $x_{1}y_{1}$, while the Add\_b blocks compute $x_{0}+x_{1}$ and $y_{0}+y_{1}$. Then, the final result is computed by performing one more multiplication followed by appropriate additions, subtractions and shifting. These are efficiently handled by the 3:2 CSA\_b Compressor and the MulAdd\_b modules.
Similarly, the multiplier inside the MulAdd\_b module can also be further decomposed into smaller units.
In our FFAU, the Mul256 module is implemented following this approach with $b = 128$, while the Mul128 module inside it is again implemented similarly with $b = 64$, thus creating 2 levels of Karatsuba multiplication. 

\subsubsection{Unified Reduction Block}

The prime $2^{255} - 19$ in Curve25519 is a pseudo-Mersenne prime which allows fast reduction of each 512-bit product using shifts and additions / subtractions \cite{salarifard_efficient_2019}.
The prime $2^{448} - 2^{224} - 1$ in Curve448 is a Solinas prime which also allows fast reduction of the 896-bit product using shifts and additions / subtractions \cite{awaludin_high_2022}.
The reduction unit can perform 4 reductions modulo \(2^{255}-19\) at once for Curve25519, simultaneously reducing all 4 of the 512-bit products computed by the 4 multipliers in the FFAU. The adders required for this pseudo-Mersenne prime reduction are re-used for reducing the 896-bit product modulo \(2^{448} - 2^{224} - 1\) for Curve448.  

\subsubsection{Modular Inversion}

Fermat's Little Theorem \cite{menezes_handbook_2018} has been employed to perform modular inversion at the end of the ECSM computation for both Curve25519 and Curve448. This requires iteratively computing 265 and 462 modular multiplications, respectively, for Curve25519 and Curve448. These two-operand multiplications are executed in the FFAU by computing \(A \times C\) as \((A + 0) \times (C + 0)\). The corresponding control signals and register addresses are also stored in LUTs similar to the LADDER computations.



\subsection{Side-Channel Countermeasures}

Side-channel attacks \cite{verbauwhede_embedded_2006} exploit physical leakages such as timing, power consumption and electromagnetic emissions to extract secret information from software and hardware implementations of cryptographic algorithms. Ensuring side-channel resilience is crucial for maintaining the security and integrity of cryptographic operations.
Elliptic curves like Curve25519 and Curve448 are designed to have constant-time ECSM computation independent of the secret scalar, thus reducing the risk of timing attacks \cite{joye_ladder_2002}. This automatically prevents SPA attacks as well.
In order to further enhance side-channel resilience by preventing DPA attacks, randomized projective coordinates \cite{fan_ecc_2010} are used to represent elliptic curve points during the ECSM computation.
The randomization process involves transforming a point ($X$, $Y$, $Z$) in projective coordinates to ($\lambda X$, $\lambda Y$, $\lambda Z$), where $\lambda$ is a pseudo-random non-zero scalar. This helps to obscure the correlation between physical measurements and the internal arithmetic operations.

In our accelerator, the scalar $\lambda$ is generated using a PRNG and it is multiplied with the input coordinates in the FFAU before the ECSM computations begin. The pseudo-random scalar is 255-bit for Curve25519 and 448-bit for Curve448. This additional security feature has negligible impact on performance as it requires only few additional clock cycles.

The PRNG contains a light-weight Trivium stream cipher \cite{preneel_trivium_2006} which can generate 64 cryptographically secure pseudo-random bits per clock cycle. These 64-bit outputs are concatenated over multiple cycles to obtain the 255-bit and 448-bit wide pseudo-random scalars. The Trivium core operates with a 288-bit internal state initialized using an 80-bit key and an 80-bit initialization vector (IV).

\section{Implementation Results}
\label{sec:implementation}

\begin{figure}[!b]
\centering
\includegraphics[width=3.4in]{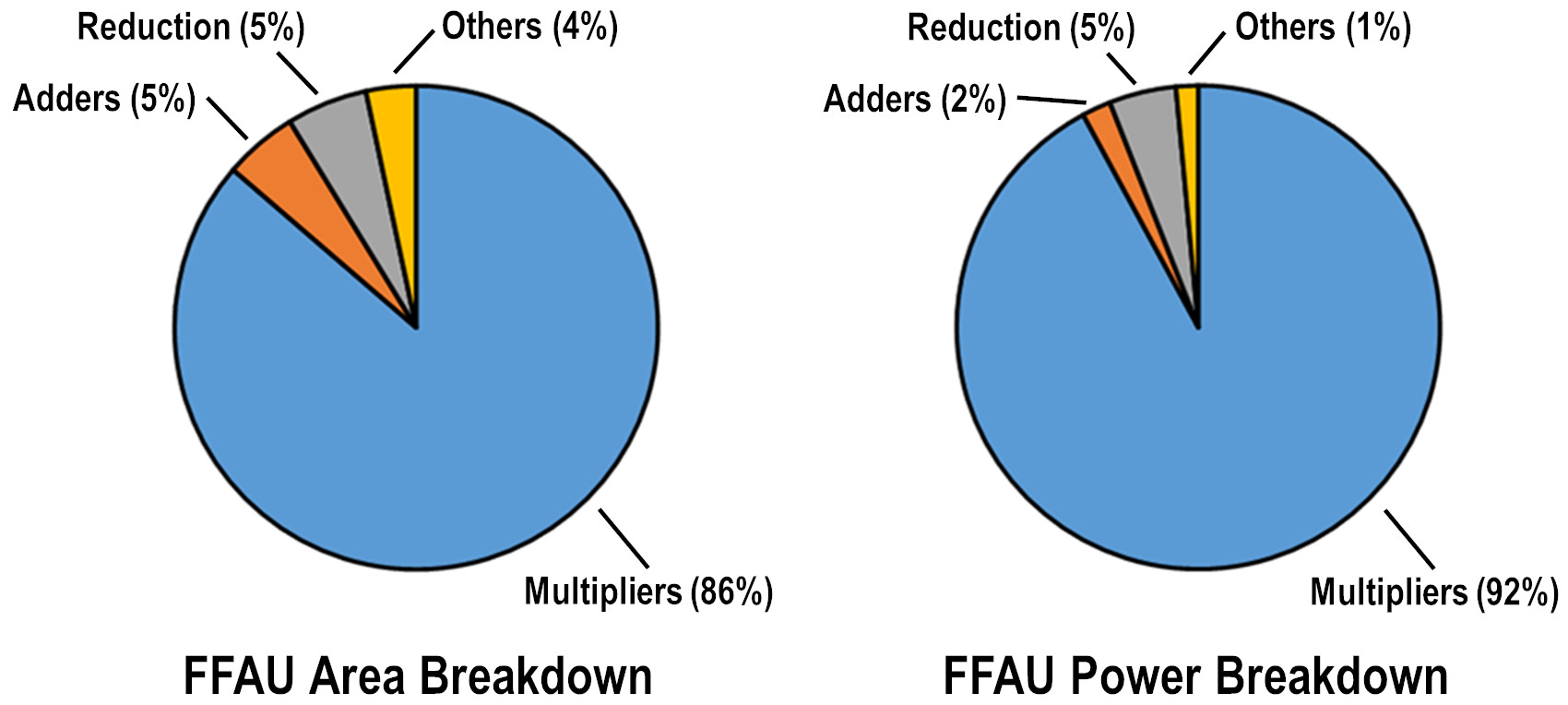}
\caption{Area and power breakdown of the FFAU.}
\label{fig:FFAU_area_power}
\end{figure}

We design our accelerator using Verilog HDL and verify its functionality with Cadence Incisive v15.20-s086. We implement the accelerator in a commercial 28nm ASIC technology and obtain post-synthesis simulation results with Cadence Genus v21.18-s082\_1 and Cadence Joules v21.18-s002\_1.
Our synthesized ASIC implementation operates at a maximum frequency of 100~MHz, occupies 1096~kGE (gate equivalent) area and consumes around 69~mW power at 0.9~V supply voltage under typical operating conditions. The FFAU consumes 93\% of the area and 97\% of total power in the accelerator. The registers consumes 6\% of the area and 2\% of total power, while the remaining 1\% is due to the PRNG and control logic. Area and power breakdown of the FFAU in terms of multipliers (4 $\times$ Mul256), adders (8 $\times$ Add255 and 2 $\times$ Add193), modular reduction and other control logic is shown in Fig. \ref{fig:FFAU_area_power}. Clearly, the 256-bit multipliers account for majority of the area and power consumption within the FFAU. The critical path of the accelerator also lies in the complex modular arithmetic circuitry (multiplications, additions / subtractions and modular reduction) present inside the FFAU.

\begin{table*}[!t]
\renewcommand{\arraystretch}{1.15}
\caption{Comparison with State-of-the-Art High-Performance ECSM Hardware Accelerators}
\label{table:comparison}
\centering
\begin{tabular}{|l|c|c|c|c|c|c|c|c|c|c|}
\hline
\textbf{Design} & \textbf{Implementation} & \textbf{Supported} & \textbf{Voltage} & \textbf{Freq.} & \textbf{Area} & \textbf{Power} & \textbf{ECSM} & \textbf{ECSM} & \textbf{SPA} & \textbf{DPA} \\
& \textbf{Platform} & \textbf{Curve(s)} & \textbf{(V)} & \textbf{(MHz)} & & \textbf{(mW)} & \textbf{Latency} & \textbf{Energy} & \textbf{CM $^1$} & \textbf{CM $^1$} \\
\hline
\hline
\textbf{This} & \multirow{2}{*}{\textbf{28nm ASIC $^2$}} & \textbf{Curve25519} & \multirow{2}{*}{\textbf{0.9}} & \multirow{2}{*}{\textbf{100}} & \multirow{2}{*}{\textbf{1096~kGE}} & \multirow{2}{*}{\textbf{69}} & \textbf{10.38~$\mu$s} & \textbf{0.72~$\mu$J} & \multirow{2}{*}{\textbf{Yes}} & \multirow{2}{*}{\textbf{Yes}} \\ \cline{3-3} \cline{8-9}
\textbf{Work} & & \textbf{Curve448} & & & & & \textbf{54.01~$\mu$s} & \textbf{3.73~$\mu$J} & & \\
\hline
\cite{salarifard_efficient_2019} & 45nm ASIC $^2$ & Curve25519 & 1.1 & 102 & 541~kGE & - & 52~$\mu$s & - & Yes & Yes \\
\hline
\cite{do_multi_2023} & 180nm ASIC $^3$ & Curve25519 & 1.8 & 102 & 377~kGE & 627 & 8.54~ms & 5.35~mJ & Yes & - \\
\hline
\multirow{2}{*}{\cite{niasar_fast_2020}} & Zynq 7020 & \multirow{2}{*}{Curve25519} & \multirow{2}{*}{-} & \multirow{2}{*}{60} & 6,183 Logic Slices & \multirow{2}{*}{-} & \multirow{2}{*}{103~$\mu$s} & \multirow{2}{*}{-} & \multirow{2}{*}{Yes} & \multirow{2}{*}{Yes} \\
& FPGA & & & & + 81 DSPs + 0.5 BRAMs & & & & & \\
\hline
\multirow{2}{*}{\cite{wu_high_2022}} & Zynq 7000 & \multirow{2}{*}{Curve25519} & \multirow{2}{*}{-} & \multirow{2}{*}{204} & 5,403 Logic Slices & \multirow{2}{*}{-} & \multirow{2}{*}{14~$\mu$s} & \multirow{2}{*}{-} & \multirow{2}{*}{Yes} & \multirow{2}{*}{-} \\
& Series FPGA & & & & + 128 DSPs + 24 BRAMs & & & & & \\
\hline
\multirow{2}{*}{\cite{sasdrich_cryptography_2017}} & Zynq 7020 & \multirow{2}{*}{Curve448} & \multirow{2}{*}{-} & \multirow{2}{*}{335} & 1,648 Logic Slices & \multirow{2}{*}{-} & \multirow{2}{*}{1.41~ms} & \multirow{2}{*}{-} & \multirow{2}{*}{Yes} & \multirow{2}{*}{Yes} \\
& FPGA & & & & + 35 DSPs + 14 BRAMs & & & & & \\
\hline
\multirow{2}{*}{\cite{niasar_optimized_2020}} & Zynq 7020 & \multirow{2}{*}{Curve448} & \multirow{2}{*}{-} & \multirow{2}{*}{95} & 4,424 Logic Slices & \multirow{2}{*}{-} & \multirow{2}{*}{1.4~ms} & \multirow{2}{*}{-} & \multirow{2}{*}{Yes} & \multirow{2}{*}{Yes} \\
& FPGA & & & & + 81 DSPs & & & & & \\
\hline
\multirow{2}{*}{\cite{awaludin_high_2022}} & Virtex-7 & \multirow{2}{*}{Curve448} & \multirow{2}{*}{-} & \multirow{2}{*}{245} & 7,666 Logic Slices & \multirow{2}{*}{-} & \multirow{2}{*}{200~$\mu$s} & \multirow{2}{*}{-} & \multirow{2}{*}{Yes} & \multirow{2}{*}{Yes} \\
& FPGA & & & & + 88 DSPs & & & & & \\
\hline
\cite{awano_fourq_2019} & 65nm ASIC $^4$ & FourQ & 1.2 & 250 & 1400~kGE & 394 & 10.1~$\mu$s & 3.98~$\mu$J & Yes & - \\
\hline
\multirow{2}{*}{\cite{jarvinen_fourq_2016}} & Zynq 7020 & \multirow{2}{*}{FourQ} & \multirow{2}{*}{-} & \multirow{2}{*}{190} & 1,691 Logic Slices & \multirow{2}{*}{-} & \multirow{2}{*}{157~$\mu$s} & \multirow{2}{*}{-} & \multirow{2}{*}{Yes} & \multirow{2}{*}{-} \\
& FPGA & & & & + 27 DSPs + 10 BRAMs & & & & & \\
\hline
\cite{knevzevic_low_2016} & 45nm ASIC $^2$ & NIST P-256 & 1.1 & 295 & 1034~kGE & - & 37~$\mu$s & - & Yes & - \\
\hline
\cite{tamura_ecdsa_2016} & 65nm ASIC $^4$ & Any & 1.2 & 105 & 1.92~mm$^2$ & 43 & 325~$\mu$s & 13.9~$\mu$J & Yes & - \\
\hline
\cite{tamura_montgomery_2016} & 65nm ASIC $^3$ & Any & 1.2 & 105 & 2490~kGE & 178 & 60~$\mu$s & 10.7~$\mu$J & Yes & - \\
\hline
\multicolumn{11}{l}{\footnotesize{For all previous work, the fastest implementations with side-channel (SPA and/or DPA) countermeasures are considered for fair comparison.}} \\
\multicolumn{11}{l}{\footnotesize{$^1$ CM: Countermeasures $\,\,\,^2$ post-synthesis simulation results $\,\,\,^3$ post-layout simulation results $\,\,\,^4$ post-silicon measurement results}}
\end{tabular}
\end{table*}

For Curve25519, each ECSM computation takes 1,032 and 1,038 clock cycles respectively without and with randomized projective coordinate DPA countermeasures. This corresponds to ECSM latency of 10.32~$\mu$s and 10.38~$\mu$s respectively. The corresponding energy consumption per ECSM operation are 0.71~$\mu$J and 0.72~$\mu$J respectively.
For Curve448, each ECSM computation takes 4,944 and 5,401 clock cycles respectively without and with randomized projective coordinate DPA countermeasures. This corresponds to ECSM latency of 49.44~$\mu$s and 54.01~$\mu$s respectively. The corresponding energy consumption per ECSM operation are 3.41~$\mu$J and 3.73~$\mu$J respectively.
The ECSM computation times achieved by our design for both curves are well within the requirements of latency-critical industrial communication protocols such as IEC 61850 \cite{hohlbaum_practical_2010}.
Also, the DPA countermeasure, that is, randomization of projective coordinates, has almost no impact on overall performance and energy-efficiency.

Table \ref{table:comparison} compares our design with previous work on high-performance ECSM hardware accelerators implemented in FPGA and ASIC. Our proposed accelerator not only supports two curves at different security levels in the same hardware but also incorporates SPA and DPA countermeasures.
Our design achieves better performance and lower energy consumption compared to previous Curve25519 and Curve448 accelerators \cite{salarifard_efficient_2019, do_multi_2023, niasar_fast_2020, wu_high_2022, sasdrich_cryptography_2017, niasar_optimized_2020, awaludin_high_2022}.
Compared to previous work on ECC accelerators for other curves at 128-bit security level such as FourQ and NIST P-256 \cite{awano_fourq_2019, jarvinen_fourq_2016, knevzevic_low_2016, tamura_ecdsa_2016, tamura_montgomery_2016}, our design achieves better or similar performance and energy consumption while supporting a higher security level curve as well as stronger side-channel countermeasures.
Compared to previous work on low-power ASIC implementations of ECC hardware accelerators \cite{hutter_nacl_2015, banerjee_jssc_2019}, our design has lower energy consumption but much larger area due to the high performance requirement.

\section{Conclusions and Future Work}
\label{sec:conclusion}

In this work, we have presented a high-performance unified hardware accelerator for elliptic curve scalar multiplication (ECSM) over NIST standard Montgomery curves Curve25519 and Curve448.
We implement an efficient finite field arithmetic unit (FFAU) with four 256-bit 2-level Karatsuba multipliers to enable parallel processing of arithmetic operations.
We restructure the sequence of operations in the Montgomery Ladder for faster computation and store the corresponding instructions in lookup tables for efficient control.
Our proposed design strategy facilitates the concurrent execution of up to four 255-bit arithmetic operations during Curve25519 ECSM computation. The same circuitry is re-used for the execution of one 448-bit arithmetic operations during Curve448 ECSM computation.
We implement a unified modular reduction block which enables fast reduction using special mathematical properties of the pseudo-Mersenne and Solinas primes in Curve25519 and Curve448 respectively.
Our implementation is constant-time by design and the Montgomery Ladder ensures inherent resilience against SPA attacks. Using a Trivium-based PRNG, we also incorporate the randomized projective coordinate countermeasure to prevent DPA attacks with negligible impact on performance. 
Our ASIC implementation achieves record performance and energy of 10.38~$\mu$s / 54.01~$\mu$s and 0.72~$\mu$J / 3.73 $\mu$J respectively for Curve25519 / Curve448. This is significantly better than state-of-the-art, which makes our design particularly attractive for latency-critical applications.
Our proposed architecture will benefit electronic systems which need to support elliptic curve cryptography at different security levels based on the requirements of the target applications and can easily switch between Curve25519 and Curve448 to achieve security-versus-efficiency trade-offs.

As future work, our proposed hardware architecture can be extended to incorporate additional side-channel countermeasures. With minor modifications to its controller, the design can also be extended to support Montgomery Ladder ECSM computation with other curves, underscoring its versatility.


\section*{Acknowledgment}

The research presented in this work was conducted as part of the project ``Efficient and Side-Channel-Resilient Implementation of Cryptographic Algorithms and Security Protocols in Embedded Systems for Power Grid Applications'' funded by the POWERGRID Center of Excellence in Cyber Security (PGCoE), Indian Institute of Science, Bangalore. 

\bibliographystyle{IEEEtran}
\bibliography{references}

\end{document}